# Valley-locked Optical Spin Merons in Valley Photonic Crystal Waveguides

Lvjin He[1], Shanshan Chen[1], Ziyang Chen[1], Lan Zhang[1], Lipeng Wan[1], Weimin Deng[1,*], and Tianbao Yu[1,*]

[1] *School of Physics and Materials Science, Jiangxi Provincial Key Laboratory of Photodetectors, Nanchang University, Nanchang 330031, China.*

[*] Corresponding author: dengweimin@ncu.edu.cn; yutianbao@ncu.edu.cn;

## Abstract

Optical skyrmionic textures, including Néel-type skyrmions and merons, have attracted significant attention across diverse physical systems for their promising applications in ultra-precise metrology, optical information processing, and quantum technologies. However, the lack of an effective approach for their on-chip directional transport and manipulation impedes their applications in photonic integrated devices. Here, we numerically demonstrate a photonic platform that utilizes topologically protected valley edge state to achieve robust on-chip directional transport of optical spin merons. These merons originate from spin-orbit coupling within the evanescent field at the valley photonic crystal surface and exist as eigenstates of the topologically protected edge state, ensuring their robust propagation even in the interface with defects. Leveraging the valley degree of freedom of topological edge states, we further achieve valley-locked spin merons, enabling flexible control over the polarity of spin merons. By endowing spin merons with topological protection in momentum space, our work provides an approach for robust on-chip transport and manipulation of spin merons, thereby paving the way for expanding their application in photonic systems.

## Introduction

Topological spin textures, characterized by nontrivial spin configurations, have aroused significant attention in diverse research fields. Skyrmionic textures are initially explored in nuclear physics[1], and later extended to condensed matter systems, including Bose-Einstein condensates[2] and magnetic materials[3–6]. Their configurations correspond to continuous vector fields that achieve complete or half coverage of a parametric sphere. The topology of such configurations is characterized by a quantized topological invariant, i.e. skyrmion number, which directly confers exceptional stability against external perturbations.

Inspired by their unique topology, the concepts of skyrmionic textures such as skyrmions and merons have been successfully extended to classical wave systems, encompassing water waves[7,8], acoustic waves[9–11], and optical waves[12,13]. In the optical doamin, these topologically nontrivial configurations can be constructed by various vectors of light, primarily including the electromagnetic field vectors[14–20], spin vectors[21–27], and Stokes vectors[28–31]. So far, skyrmionic textures have been observed across diverse physical systems, such as evanescent electromagnetic fields[14,21], structured light beams[32,33], bound states in the continuum[30] and even intricate spatiotemporal optical fields[17,24,34,35]. With topological stability and configurable vectorial textures, optical skyrmions have enabled significant applications in sensing[36], metrology[37], optical information processing[38–40], and quantum information[41,42].

Besides, the topological structure of optical skyrmionic textures ensures their stability against perturbations, making them the ideal candidate for robust information carriers in on-chip data transmission and processing. So far, the on-chip generation of optical skyrmionic textures is mainly realized by the tailored interference fields of surface plasmon polaritons[14,16,18,23,26], which suffers from absorption loss of metals and cannot support the directional transport over long distances. Thus, it remains challenging to achieve robust on-chip propagation and manipulation of optical skyrmions, which severely limits their potential applications.

Topological photonic crystals (TPCs) support robust topological edge states (TESs)

that facilitate backscattering-immune and defect-tolerant robust transmission, offering significant advantages for photonic integrated devices[43–53]. TESs can be realized in all-dielectric valley photonic crystals (VPCs) based on the photonic analogue of quantum valley Hall effect[49–51,54,52,53], and the existence of TESs can be predicted by the valley Chern number, a topological invariant defined in momentum-space. VPCs offer a promising platform for on-chip integration because they require neither magneto-optical materials nor pseudo-spin engineering, and their TESs are located below the light cone, thus avoiding radiation loss. Thus, an open question is whether we can realize robust on-chip directional transport of optical skyrmions by TESs.

In this work, we numerically demonstrate the robust on-chip transport of optical spin merons via a VPC waveguide (VPCW). As shown in Fig. 1, we reveal that the eigenmodes of the TESs carry spin merons, enabling directional transport of spin merons. These nontrivial spin textures arise from the spin-orbit coupling (SOC) within the evanescent vortex field on the surface of VPCW. On each side of the edge, the eigenmodes have opposite local orbital angular momentum (OAM), thus carrying spin merons with opposite polarity. By exploiting the valley degree of freedom in VPCs, we further realize valley-locked spin merons, indicating that the sign of skyrmion number is locked to the propagation direction of the TES. Benefitting from the topological protection of TESs, the generated meron flow exhibits good robustness, with inherent tolerance to sharp bends and structural defects. Our work paves the way to the experimental realization of on-chip directional transmission and manipulation of spin merons.

## Results

The VPCW designed in this work is constructed by two spatially inverted VPCs (VPC1 and VPC2), as schematically illustrated in the left panel of Fig. 2(a). Both VPC1 and VPC2 have the same lattice constant $a$ = 20 μm. Their unit cells consist of equilateral triangular germanium rods (n = 4) with a height of $h$ = 3 μm and a side length of $l$ = 17 μm, which are periodically arranged on a perfect electric conductor (PEC) substrate. When the rotation angle of dielectric rods in VPC1 and VPC2 is $\alpha = 0^{\circ}$, the $C_{3v}$ point group symmetry at K (and K′) ensures the degeneracy of the transverse magnetic-like (TM-like) mode at K and K' point, as shown by the gray bands in the middle panel of Fig. 2(a). When rotating the triangular dielectric rods in VPC1 (VPC2) by 30 (-30) degrees, the symmetry of both VPC1 and VPC2 reduces from $C_{3v}$ to $C_3$, which breaks the degeneracy at K/K′ point, and consequently induces the formation of a complete bandgap (CBG). As inversion symmetry partners, VPC1 and VPC2 exhibit an identical band structure, as shown by the black bands in the middle panel of Fig. 2(a). By integrating the Berry curvature of the lower band around K and K′ valleys, i.e., $C_{K(K')} = \int_{K(K')} \Omega_n(k_x, k_y) d^2 k$, valley Chern numbers ($C_V = C_K - C_{K'}$) of VPC1 (VPC2) is $-1\,(1)$, which confirms their nontrivial topology and guarantees the existence of TESs at the VPC1/VPC2 interface. Figure 2(b) shows the schematic of the VPCW and its projected bands, where black curves and grey areas represent the bands of TES and bulk states, respectively. The red dashed lines represent light line in air. The TES at K (K') valley is highlighted by blue (red) dot. The $|E_z|$ distribution of the TES at K′ valley is shown in the left side of projected bands.

The phase distributions of $E_z$ and $H_z$ for TES at K valley are depicted in Fig. 2(c). The black dashed line highlights the position of the interface. The plus and minus symbols represent a phase vortex with a topological charge of 1 and -1, respectively. In the inner region of the selected hexagonal regions, the $E_z$ component of TES at the K valley carries phase vortex with opposite topological charge in VPC1 region ($l = 1$) and VPC2 region ($l = -1$). In addition, near the corners of the hexagon regions, there are also phase singularities, and the topological charge of the associated phase vortexes are opposite to those inside the hexagon. For the phase of $H_z$ component, phase vortexes also exist in both the inner region and corners of the hexagon, but their positions are different from that of $E_z$ component, as shown in the right panel of Fig. 2(c). Compared with the TES at K valley, due to time-reversal symmetry (TRS), TES at K′ valley exhibits similar eigenfield distributions but opposite phase vortexes in the same hexagonal regions, as shown in Fig. 2(e).

For the electromagnetic fields above the surface of the VPCW, their spin angular momentum (SAM) density can be calculated by

$$\boldsymbol{S} = \boldsymbol{S_E} + \boldsymbol{S_H} = \frac{1}{4\omega}(Im(\varepsilon_0 \boldsymbol{E}^* \times \boldsymbol{E}) + Im(\mu_0 \boldsymbol{H}^* \times \boldsymbol{H})) \quad (1)$$

This expression reveals that SAM can be expressed as the sum of the electric SAM ($S_E$) and magnetic SAM ($S_H$). Figure 2(d) shows the distributions of normalized electric-part, magnetic-part and total $S_z$ component for the TESs at K valley, plotted at a distance of 2 μm from the VPCW surface. For the electric-part $S_z$ component ($S_{z,E}$) in VPC1 region, sgn($S_{z,E}$) is positive at the position of phase vortex inside the selected hexagonal region while negative at the position of phase vortexes at the corners. This phenomenon can be inferred by the distribution of phase vortexes. Note that $S_{z,E}$ is associated with the polarization ellipticity of in-plane electric fields and its direction is determined by the sign of phase vortex's topological charge[55]. Therefore, the phase vortexes with opposite topological charge will induce $S_{z,E}$ components with opposite directions. Since $E_z$ vanishes at the position of $E_z$'s phase singularity, only the $S_{z,E}$ component is non-zero. As a result, $\mathbf{S}_\mathrm{E}$ is usually completely upward (downward) at the position where the phase vortex possesses positive (negative) topological charge. From the phase vortex inside the selected hexagonal region to the surrounding three phase vortexes with opposite topological charge, $S_z$ undergoes a reversal of direction, which is conducive to the formation of merons. Similar analysis can be applied to magnetic-part $S_z$ component ($S_{z,H}$). A little difference is that the intensity of $H_y$ component is very small at the positions of $H_z$'s phase vortexes near the corners of the selected hexagonal region, leading to a very small $S_{z,H}$ (see more details in supplementary note1). By considering the contribution from both $S_{z,E}$ and $S_{z,H}$, the normalized $S_z$ experiences a direction inversion from the center to three corner regions of the selected hexagonal regions in both VPC1 and VPC2 region, as shown in Fig. 2(d).

Apart from $S_z$ component, $S_x$ and $S_y$ components are also necessary for the generation of a spin meron. Here, we utilize the property of the evanescent wave.

Because the TES considered here lies below the light cone (Fig. 2b), it does not radiate into free space and decays evanescently along the z direction. This introduces a $\pi/2$ phase difference between the in-plane and out-of-plane field components, which gives rise to in-plane $S_x$ and $S_y$ components[56]. Consequently, the evanescent fields on the VPCW surface, combined with the phase vortexes, leads to complete three-dimensional SAM components. The right panel of Fig. 2(d) shows the distribution of three-dimensional spin vectors and in-plane azimuthal angle of SAM (denoted as $\mathrm{atan}(S_y/S_x)$) within the selected hexagonal regions. For the region enclosed by a black dashed curve inside each hexagon, the spin vectors form a meron. The azimuthal angle singularity highlighted by a black dot indicates the in-plane spin vectors possess a non-zero vorticity. To quantify the topological nontriviality of these spin textures, we calculate the skyrmion number $n_{sk}$ using the following equation[12]:

$$n_{sk} = \frac{1}{4\pi}\iint \hat{\boldsymbol{S}} \cdot \left(\frac{\partial \hat{S}}{\partial x} \times \frac{\partial \hat{S}}{\partial y}\right) dxdy, \tag{2}$$

where $\hat{\boldsymbol{S}}$ is the normalized SAM density. For the spin texture in the VPC1 region (meron 1), the calculated skyrmion number is $n_{sk} = -0.498$. Here, a spin texture with $|n_{\mathrm{sk}}| > 0.49$ is identified as a meron. The right panel of Fig. 2(d) also shows the spin texture in VPC2 region (meron 2), the direction of $S_z$ is opposite to that of meron1 and the calculated skyrmion number is $n_{sk} = 0.499$. Thus, the TES at K valley carries spin merons in both VPC1 and VPC2 regions. Note that the phase vortexes' topological charges carried by the TES at K' valley are reversed compared to that carried by the TES at K valley for the same hexagon region. Therefore, compared with the TESs at the K valley, spin textures of the TESs at the K' valley have opposite polarity and skyrmion number (Fig. 2(f)), meaning that the sign of skyrmion number is locked to valley degree of freedom. This feature enables us to flexibly modulate the polarity of merons by selectively exciting the TESs at K or K′ valley

Since we have revealed the existence of spin merons in the eigenmode of TESs, one can realize on-chip directional transport of spin merons in the VPCW. To demonstrate the transport of the merons, we excite the TES in the VPCW. The left panel of Fig. 3(a) depicts the $|E_z|$ distribution of the excited TES at the K valley, and the right panel shows the normalized $S_z$ distribution in the region highlighted by the white dashed rectangle. One can see spin merons periodically distribute around the interface, indicating these nontrivial spin textures can propagate along with the TES. To evaluate the bandwidth of spin merons, we sweep the frequency of the excitation source and calculate the skyrmion numbers in the two hexagonal regions, as shown in the left panels of Fig. 3(b) and (c). The results indicate that the directional transport of spin merons is not limited to the TESs at K valley, but remains effective over a certain frequency range. Right panels in Fig. 3(b) and (c) highlights spin vectors and the in-plane azimuthal angle of SAM for meron1 and meron2, respectively. For the TES at K' valley, spin merons can also propagate along the interface but towards the opposite direction (Fig. 3(d)). This behavior is also observed in a certain frequency range, according to the calculated skyrmion numbers in Figs. 3(e) and (f). Once the TES's propagation direction reverses, the sign of the skyrmion number will also reverse, which is consist with the results shown in Fig. 2(d) and (f).

Because the transport of TESs is robust to sharp bends and some defects, one may expect that the transport of spin merons will also possess robustness. To verify the robustness of the spin merons propagating along the VPCW, we investigate their transport in waveguides with defects. Figure 4(a) depicts the $|E_z|$ distribution the TES at $K'$ valley in the VPCW with defect I. Here, defect I is created by twisting two dielectric rods near the interface of the VPCW by 30 degrees, as highlighted by green triangles in the inset of Fig. 4(a). The $|E_z|$ distribution clearly shows that the transport of the TES is nearly unaffected by the defects. Figures 4(b) and 4(c) depict the normalized $S_z$ distribution for the region with defect I and without defect, respectively. The right panels present the distribution of spin vectors and the in-plane azimuthal angle of the SAM within selected hexagons. Although the spin textures near the defects become deformed, merons still exist and retain their intrinsic skyrmion numbers. Furthermore, away from the defect region, the merons recover their original shape (Fig. 4(c)), which is similar to the $S_z$ distribution in the right panel of Fig. 3(d). As shown in Fig. 4(d), we further consider the case of four rotated dielectric rods near the interface of VPCW (defect II). The unidirectional transport of the TES remains almost unaffected. The results in Figs. 4(e) and 4(f) show that merons still exist at the region with more defects, which demonstrates the nontrivial topology and robustness of merons. Above results prove that the transport of spin merons in the VPCW possesses excellent robustness against structural defects induced by rotating dielectric rods. Besides, we further study the transport robustness of spin meron in the VPCW with a Z-shaped bend. As shown in Fig. 4(g), benefiting from the backscattering-immune property, the TESs propagates smoothly through the sharp bend region. As shown in Fig. 4(h), the spin texture undergoes significant deformation at the corner of Z-shaped bend, resulting in the absence of merons at the selected hexagon region in VPC2. However, away from the corner, merons reappear and gradually recover their shape (Fig. 4(i)), see more details in supplementary note2.

In conclusion, we numerically demonstrate a VPCW platform for on-chip generation and robust directional transport of optical spin merons. Valley-locked spin merons are formed via SOC within the evanescent field of TESs, enabling flexible control over the polarity of spin merons by selectively exciting the TES at K or K' valley. The generated spin merons show excellent propagation robustness against structural defects. For the VPCW with a Z-shaped bend, although the spin meron undergoes significant deformation at the corner, it shows the self-healing ability and reappears away from the corner. This work endows spin merons with topological protection in momentum space, overcoming the limitations of conventional schemes in realizing robust on-chip transport and manipulation of spin merons. The proposed VPCW provides a good platform for the design of skyrmion-based photonic devices and has potential applications in robust on-chip data transmission and processing.

**Acknowledgements**

This work is supported by the National Natural Science Foundation of China (12564052, 62305146); Training Program for Academic and Technical Leaders of Major Disciplines in Jiangxi Province (20243BCE51163); Innovation Leader of the Ganpo Talent Program; Natural Science Foundation of Jiangxi Province (20242BAB20023, 20244BDF60006); Jiangxi Provincial Key Laboratory of Photodetectors (2024SSY03041).

**Conflict of interest**

The authors declare no competing interests.

## References

[1] T. H. R. Skyrme, A non-linear field theory Proc. R. Soc. Lond. Math. Phys. Sci. **260**, 127 (1961).

[2] U. Al Khawaja and H. Stoof, Skyrmions in a ferromagnetic Bose–Einstein condensate Nature **411**, 918 (2001).

[3] S. Mühlbauer, B. Binz, F. Jonietz, C. Pfleiderer, A. Rosch, A. Neubauer, et al., Skyrmion Lattice in a Chiral Magnet Science **323**, 915 (2009).

[4] X. Z. Yu, Y. Onose, N. Kanazawa, J. H. Park, J. H. Han, Y. Matsui, et al., Real-space observation of a two-dimensional skyrmion crystal Nature **465**, 901 (2010).

[5] X. Z. Yu, W. Koshibae, Y. Tokunaga, K. Shibata, Y. Taguchi, N. Nagaosa, et al., Transformation between meron and skyrmion topological spin textures in a chiral magnet Nature **564**, 95 (2018).

[6] A. Fert, N. Reyren, and V. Cros, Magnetic skyrmions: advances in physics and potential applications Nat. Rev. Mater. **2**, 17031 (2017).

[7] B. Wang, Z. Che, C. Cheng, C. Tong, L. Shi, Y. Shen, et al., Topological water-wave structures manipulating particles Nature **638**, 394 (2025).

[8] D. A. Smirnova, F. Nori, and K. Y. Bliokh, Water-Wave Vortices and Skyrmions Phys. Rev. Lett. **132**, 054003 (2024).

[9] H. Ge, X.-Y. Xu, L. Liu, R. Xu, Z.-K. Lin, S.-Y. Yu, et al., Observation of Acoustic Skyrmions Phys. Rev. Lett. **127**, 144502 (2021).

[10] L. Cao, S. Wan, Y. Zeng, Y. Zhu, and B. Assouar, Observation of phononic skyrmions based on hybrid spin of elastic waves Sci. Adv. **9**, eadf3652 (2023).

[11] N. Zhou, W.-N. Chen, W.-J. Sun, Z.-Q. Sheng, and H.-W. Wu, Observation of acoustic meron textures Appl. Phys. Lett. **125**, 042204 (2024).

[12] Y. Shen, Q. Zhang, P. Shi, L. Du, X. Yuan, and A. V. Zayats, Optical skyrmions and other topological quasiparticles of light Nat. Photonics **18**, 15 (2024).

[13] A. Yang, A. Kong, F. Meng, X. Chen, M. Lin, P. Shi, et al., Optical skyrmions: from fundamentals to applications J. Opt. **27**, 043002 (2025).

[14] S. Tsesses, E. Ostrovsky, K. Cohen, B. Gjonaj, N. H. Lindner, and G. Bartal, Optical skyrmion lattice in evanescent electromagnetic fields Science **361**, 993 (2018).

[15] T. J. Davis, D. Janoschka, P. Dreher, B. Frank, F.-J. Meyer Zu Heringdorf, and H. Giessen, Ultrafast vector imaging of plasmonic skyrmion dynamics with deep subwavelength resolution Science **368**, eaba6415 (2020).

[16] C. Bai, J. Chen, Y. Zhang, D. Zhang, and Q. Zhan, Dynamic tailoring of an optical skyrmion lattice in surface plasmon polaritons Opt. Express **28**, 10320 (2020).

[17] Y. Shen, Y. Hou, N. Papasimakis, and N. I. Zheludev, Supertoroidal light pulses as electromagnetic skyrmions propagating in free space Nat. Commun. **12**, 5891 (2021).

[18] Z.-L. Deng, T. Shi, A. Krasnok, X. Li, and A. Alù, Observation of localized magnetic plasmon skyrmions Nat. Commun. **13**, 8 (2022).

[19] J. Schwab, A. Neuhaus, P. Dreher, S. Tsesses, K. Cohen, F. Mangold, et al., Skyrmion bags of light in plasmonic moiré superlattices Nat. Phys. **21**, 988 (2025).

[20] L. Zhang, L. Wan, W. Deng, L. Hou, J. Qiu, Q. Zou, et al., Observation of moiré plasmonic skyrmion clusters Sci. Adv. **11**, (2025).

[21] L. Du, A. Yang, A. V. Zayats, and X. Yuan, Deep-subwavelength features of photonic skyrmions in a confined electromagnetic field with orbital angular momentum Nat. Phys. **15**, 650 (2019).

[22] Q. Zhang, Z. Xie, L. Du, P. Shi, and X. Yuan, Bloch-type photonic skyrmions in optical chiral multilayers Phys. Rev. Res. **3**, 023109 (2021).

[23] X. Lei, A. Yang, P. Shi, Z. Xie, L. Du, A. V. Zayats, et al., Photonic Spin Lattices: Symmetry Constraints for Skyrmion and Meron Topologies Phys. Rev. Lett. **127**, 237403 (2021).

[24] Y. Dai, Z. Zhou, A. Ghosh, R. S. K. Mong, A. Kubo, C.-B. Huang, et al., Plasmonic topological quasiparticle on the nanometre and femtosecond scales Nature **588**, 616 (2020).

[25] M. Lin, Q. Liu, H. Duan, L. Du, and X. Yuan, Wavelength-tuned transformation between photonic skyrmion and meron spin textures Appl. Phys. Rev. **11**, 021408 (2024).

[26] P. Shi, X. Gou, Q. Zhang, W. Wei, H. Wu, Z. Zhu, et al., Observation of strong spin-orbit

couplings in plasmonic spin-twistronics topological lattices Nat. Commun. **17**, 1905 (2026).

[27] Y. Lang, Q. Xu, F. Huang, X. Yuan, X. Jiang, X. Zhang, et al., Experimental Observation of Topological Transition in Optical Multimeron Adv. Mater. **38**, e07528 (2026).

[28] C. Guo, M. Xiao, Y. Guo, L. Yuan, and S. Fan, Meron Spin Textures in Momentum Space Phys. Rev. Lett. **124**, 106103 (2020).

[29] T. He, Y. Meng, L. Wang, H. Zhong, N. Mata-Cervera, D. Li, et al., Optical skyrmions from metafibers with subwavelength features Nat. Commun. **15**, 10141 (2024).

[30] L. Rao, J. Wang, X. Wang, S. Wu, X. Zhao, W. Liu, et al., Meron Spin Textures in Momentum Space Spawning from Bound States in the Continuum Phys. Rev. Lett. **135**, 026203 (2025).

[31] N. Mata-Cervera, D. K. Sharma, Y. Shen, R. Paniagua-Dominguez, and M. A. Porras, Skyrmionic Polarization Texture around the Phase Singularity of Optical Vortices Phys. Rev. Lett. **135**, 033805 (2025).

[32] Y. Shen, Topological bimeronic beams Opt. Lett. **46**, 3737 (2021).

[33] Y. Shen, E. C. Martínez, and C. Rosales-Guzmán, Generation of Optical Skyrmions with Tunable Topological Textures ACS Photonics **9**, 296 (2022).

[34] R. Wang, P.-Y. Bao, Z.-Q. Hu, S. Shi, B.-Z. Wang, N. I. Zheludev, et al., Observation of resilient propagation and free-space skyrmions in toroidal electromagnetic pulses Appl. Phys. Rev. **11**, 031411 (2024).

[35] Pascal Dreher, Alexander Neuhaus, David Janoschka, Alexandra Rödl, Tim Colin Meiler, Bettina Frank, et al., Spatiotemporal topology of plasmonic spin meron pairs revealed by polarimetric photo-emission microscopy Adv. Photonics **6**, 066007 (2024).

[36] A. Yang, X. Lei, P. Shi, F. Meng, M. Lin, L. Du, et al., Spin-Manipulated Photonic Skyrmion-Pair for Pico-Metric Displacement Sensing Adv. Sci. **10**, 2205249 (2023).

[37] R. Wang, P.-Y. Bao, X. Feng, J. Wu, B.-Z. Wang, and Y. Shen, Single-antenna super-resolution positioning with nonseparable toroidal pulses Commun. Phys. **7**, 356 (2024).

[38] R. Tamura, P. Kumar, A. S. Rao, K. Tsuda, F. Getzlaff, K. Miyamoto, et al., Direct imprint of optical skyrmions in azopolymers as photoinduced relief structures APL Photonics **9**, 046104 (2024).

[39] A. A. Wang, Y. Ma, Y. Zhang, Z. Zhao, Y. Cai, X. Qiu, et al., Perturbation-resilient integer arithmetic using optical skyrmions Nat. Photonics **19**, 1367 (2025).

[40] J. Jia, J. Ren, S. Zhou, Z. Zeng, H. Lin, Y. Hu, et al., Electrically tuning photonic

topological quasiparticles in synthetic two-level system Nat. Phys. **21**, 1885 (2025).

[41] P. Ornelas, I. Nape, R. de Mello Koch, and A. Forbes, Non-local skyrmions as topologically resilient quantum entangled states of light Nat. Photonics **18**, 258 (2024).

[42] J. Ma, J. Yang, S. Liu, B. Chen, X. Li, C. Song, et al., Nanophotonic quantum skyrmions enabled by semiconductor cavity quantum electrodynamics Nat. Phys. **21**, 1462 (2025).

[43] L. Lu, J. D. Joannopoulos, and M. Soljačić, Topological photonics Nat. Photonics **8**, 821 (2014).

[44] F. D. M. Haldane and S. Raghu, Possible Realization of Directional Optical Waveguides in Photonic Crystals with Broken Time-Reversal Symmetry Phys. Rev. Lett. **100**, 013904 (2008).

[45] Z. Wang, Y. Chong, J. D. Joannopoulos, and M. Soljačić, Observation of unidirectional backscattering-immune topological electromagnetic states Nature **461**, 772 (2009).

[46] A. B. Khanikaev, S. Hossein Mousavi, W.-K. Tse, M. Kargarian, A. H. MacDonald, and G. Shvets, Photonic topological insulators Nat. Mater. **12**, 233 (2013).

[47] L.-H. Wu and X. Hu, Scheme for Achieving a Topological Photonic Crystal by Using Dielectric Material Phys. Rev. Lett. **114**, 223901 (2015).

[48] F. Peng, H. Chen, L. Wan, X.-D. Chen, J. Dong, W. Deng, et al., Dual-band dislocation modes in a topological photonic crystal Photonics Res. **13**, 1554 (2025).

[49] T. Ma and G. Shvets, All-Si valley-Hall photonic topological insulator New J. Phys. **18**, 025012 (2016).

[50] J.-W. Dong, X.-D. Chen, H. Zhu, Y. Wang, and X. Zhang, Valley photonic crystals for control of spin and topology Nat. Mater. **16**, 298 (2017).

[51] X.-D. Chen, F.-L. Zhao, M. Chen, and J.-W. Dong, Valley-contrasting physics in all-dielectric photonic crystals: Orbital angular momentum and topological propagation Phys. Rev. B **96**, 020202 (2017).

[52] J.-W. Liu, F.-L. Shi, X.-T. He, G.-J. Tang, W.-J. Chen, X.-D. Chen, et al., Valley photonic crystals Adv. Phys. X **6**, 1905546 (2021).

[53] H. Xue, Y. Yang, and B. Zhang, Topological Valley Photonics: Physics and Device Applications Adv. Photonics Res. **2**, 2100013 (2021).

[54] X.-T. He, E.-T. Liang, J.-J. Yuan, H.-Y. Qiu, X.-D. Chen, F.-L. Zhao, et al., A silicon-on-insulator slab for topological valley transport Nat. Commun. **10**, 872 (2019).

[55] K. Y. Bliokh and F. Nori, Spin and orbital angular momenta of acoustic beams Phys. Rev.

B **99**, 174310 (2019).

[56] K. Y. Bliokh, D. Smirnova, and F. Nori, Quantum spin Hall effect of light Science **348**, 1448 (2015).

# Figures and figure captions

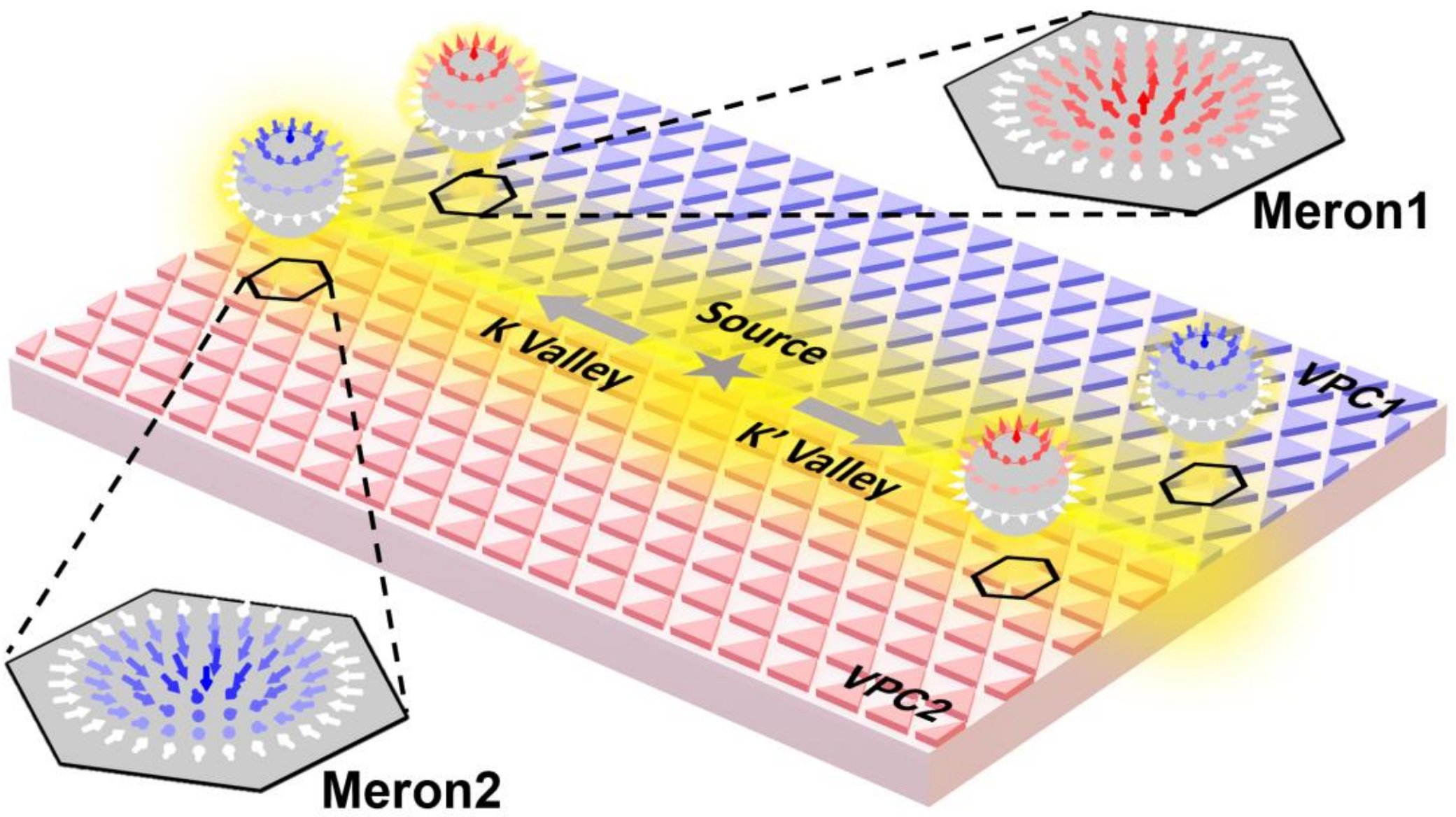

Figure 1. Schematic illustration of the formation and topologically protected transport of valley-locked spin merons in a valley photonic crystal waveguide (VPCW). The edge states at K and K′ valleys exhibit different propagation directions and opposite signs of the skyrmion number.

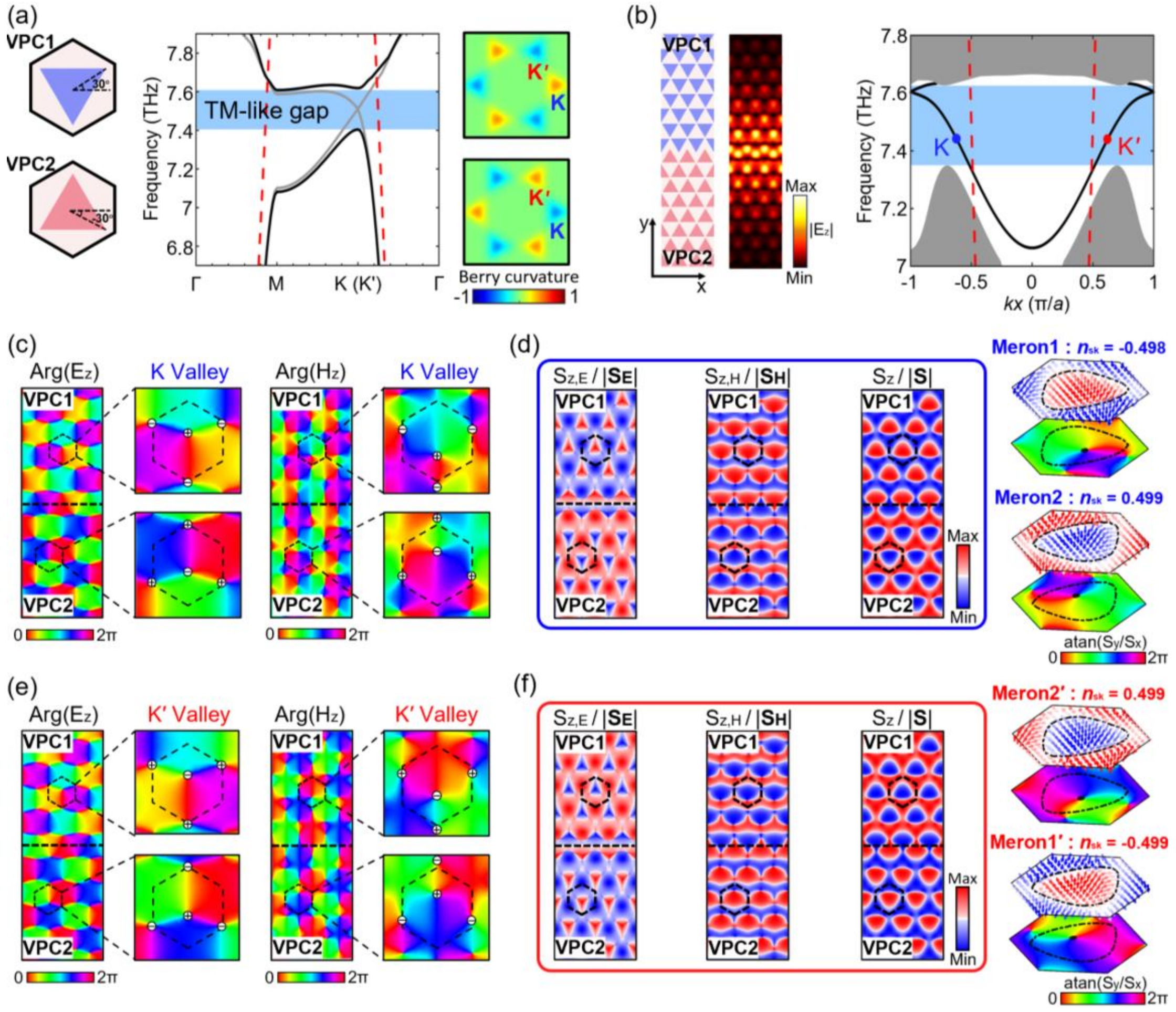

Figure 2. (a) Schematic diagram of valley photonic crystal1 (VPC1) and VPC2. The black (gray) lines represent the bulk bands of the VPC with $\alpha = 30^{\circ}$ ($\alpha = 0^{\circ}$). The light-blue area denotes the complete bandgap (CBG), and the red dashed lines plot the light cone. Right panel shows the Berry curvatures for the first band of VPC1 and VPC2. (b) Schematic diagram of the VPCW and $|E_z|$ distribution of topological edge state (TES) at K' valley. Right panel shows projected bands along $k_x$ direction. The light-blue region indicates the CBG of VPC1's bulk bands, and the grey regions represent the bulk modes. (c) Phase vortex of $E_z$ and $H_z$ for the TES at K valley. Hexagonal regions highlighted by black dashed lines denote the unit cells of VPC1 and VPC2. The plus and minus symbols represent a phase vortex with a topological charge of 1 and -1, respectively. (d) Distributions of normalized electric-part, magnetic-part and total out-of-plane spin angular momentum (SAM) for the TES at K valley. Right panels show the spin vectors and the in-plane azimuthal angle of the SAM within a unit cell of VPC1 and VPC2, where black dashed lines represent the boundary of the meron ($S_z = 0$). Black dots highlight the singularities of SAM's in-plane azimuthal angle. (e) and (f): Similar to (c) and (d), but for the TES at K' valley.

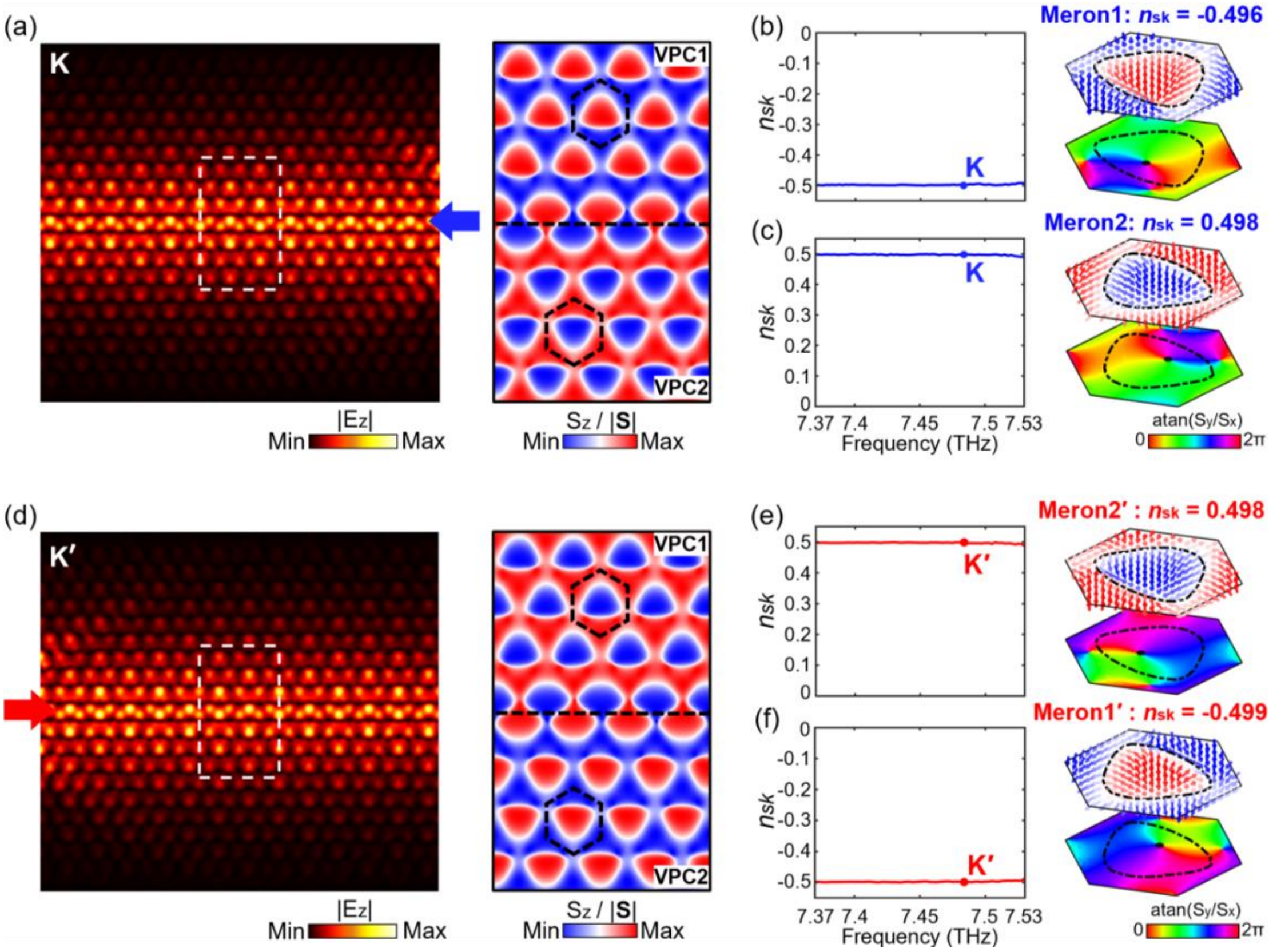

Figure 3. Directional transport of spin meron textures in the VPCW. (a) Distribution of $|E_z|$ and the normalized $S_z$ of TESs at K valley in the VPCW. The black dashed hexagons highlight the position of meron1 and meron2. (b) and (c): Skyrmion numbers as a function of frequency for meron1 and meron2. Right panels show spin vectors and the in-plane azimuthal angle of the SAM, where black dashed lines represent the boundary of the meron ($S_z = 0$). Black dots highlight the singularities of SAM's in-plane azimuthal angle. (d)-(f): Similar to (a)-(c), but for the TES at K′ valley.

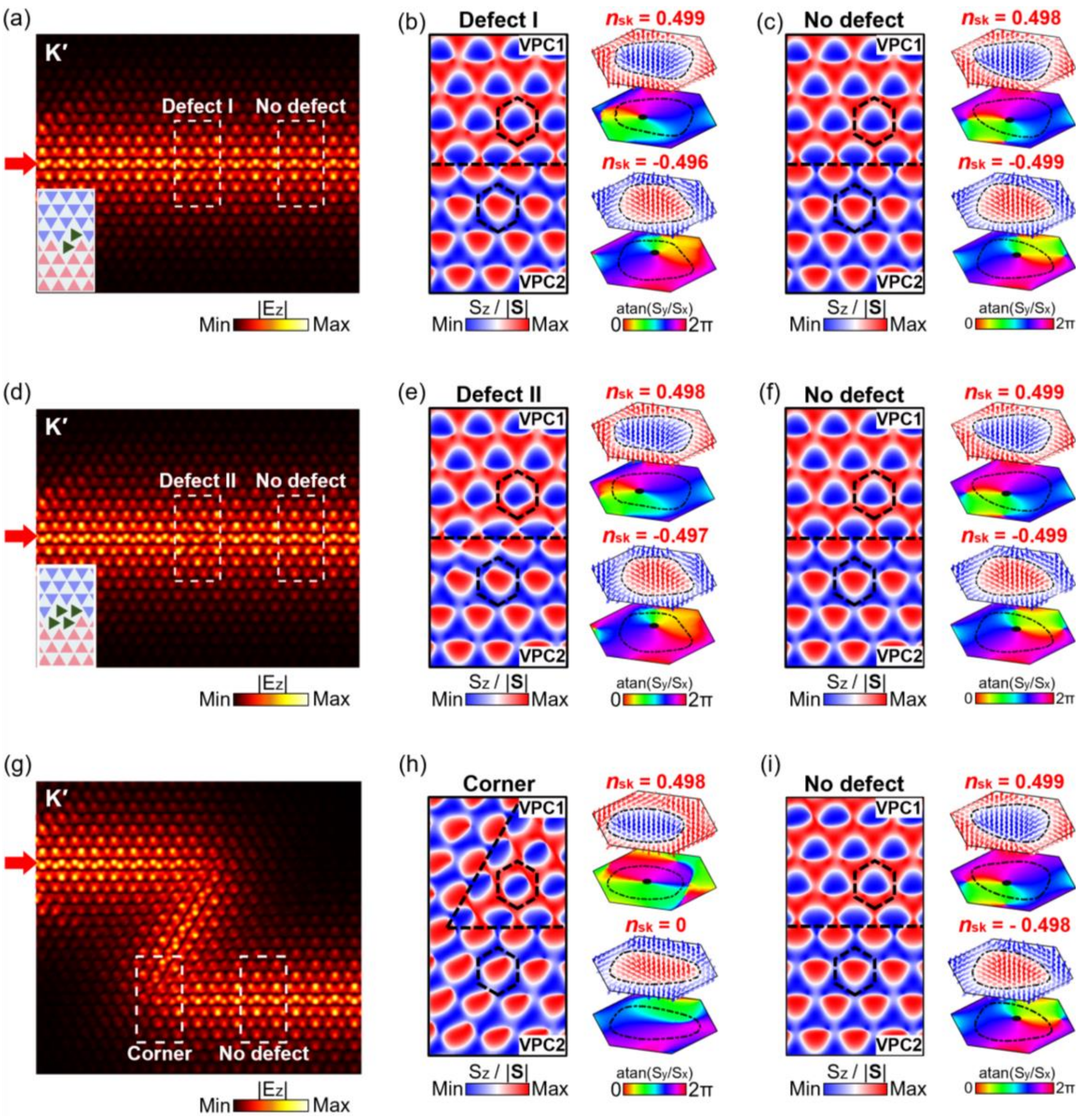

Figure 4. Robust transport of spin merons in the VPCW. (a) $|E_z|$ distribution of the TES at K′ valley in the VPCW with defect I (see the inset). (b) and (c): Normalized $S_z$ distributions for the regions with defect I and without defect in VPCW, respectively. Right panels show spin vectors and the in-plane azimuthal angle of the SAM, where black dashed lines represent the boundary of the meron ($S_z = 0$). (d-f) Similar to (a-c), but for the VPCW with defect II. (g-i) Similar to (a-c), but for the VPCW with a Z-shaped bend.